\documentclass[times, 12pt]{article}
\usepackage{times,url,graphicx}

\newtheorem{theorem}{Theorem}[section]

\newtheorem{lemma}[theorem]{Lemma}

\begin{document}
\title{Approximate Inverse Frequent Itemset Mining: Privacy,
Complexity, and Approximation\footnote{An extended abstract 
of this paper has appeared in \cite{wangwu}.}}

\author{Yongge Wang\\
SIS Department, UNC Charlotte\\
yonwang@uncc.edu\\
\and
Xintao Wu\\
CS Department, UNC Charlotte\\
xwu@uncc.edu
}

\maketitle
\thispagestyle{empty}

\begin{abstract}
In order to generate synthetic basket data sets for better benchmark
testing, it is important to integrate characteristics from real-life
databases into the synthetic basket data sets. The characteristics
that could be used for this purpose include the frequent itemsets
and association rules. The problem of generating synthetic basket
data sets from frequent itemsets is generally referred to as inverse
frequent itemset mining. In this paper, we show that the problem of
approximate inverse frequent itemset mining is {\bf NP}-complete.
Then we propose and analyze an approximate algorithm for approximate
inverse frequent itemset mining, and discuss privacy issues related
to the synthetic basket data set. In particular, we propose an
approximate algorithm to determine the privacy leakage in a
synthetic basket data set.

\end{abstract}
\section{Introduction}
Since the seminal paper \cite{seminal}, association rule and
frequent itemset mining received a lot of attention. By comparing
five well-known association rule algorithms (i.e., Apriori
\cite{Agrawal:vldb94}, Charm \cite{Zaki:kdd00}, FP-growth
\cite{Han:sigmod00}, Closet \cite{Pei:dmkd00}, and MagnumOpus
\cite{Webb:kdd00}) using three real-world data sets and the
artificial data set from IBM Almaden, Zheng et al. \cite{zheng}
found out that the algorithm performance on the artificial data sets
are very different from their performance on real-world data sets.
Thus there is a great need to use real-world data sets as
benchmarks.

However, organizations usually hesitate to provide their real-world
data sets as benchmarks due to the potential disclosure of private
information.  There have been two different approaches to this
problem. The first is to disturb the data before delivery for mining
so that real values are obscured while preserving statistics on the
collection. Some recent work
\cite{Atallah:kdex99,Dasseni:ihw01,DN03,Evfimievski:pods03,ESA+02,
Oliveira:icdm03,RH02,Saygin:sigmodr01} investigates the tradeoff
between private information leakage and accuracy of mining results.
One problem related to the perturbation based approach is that it
can not always fully preserve individual's privacy while achieving
precision of mining results.

The second approach to address this problem is to generate synthetic
basket data sets for benchmarking purpose by integrating
characteristics from real-world basket data sets that may have
influence on the software performance. The frequent sets and their
supports (defined as the number of transactions in the basket data
set that contain the items) can be considered to be a reasonable
summary of the real-world data set. As observed by Calders
\cite{calderspods}, association rules for basket data set can be
described by frequent itemsets. Thus it is sufficient to consider
frequent itemsets only. Ramesh et al. \cite{Ramesh:pods03} recently
investigated the relation between the distribution of discovered
frequent set and the performance of association rule mining. It
suggests that the performance of association rule mining method
using the original data set should be very similar to that using the
synthetic one compatible with the same frequent set mining results.

Informally speaking, in this approach, one first mines frequent
itemsets and their corresponding supports from the real-world basket
data sets. These frequent itemset support constraints are used to
generate the synthetic (mock) data set which could be used for
benchmarking. For this approach, private information should be
deleted from the frequent itemset support constraints or from the
mock database. The authors of \cite{calderspods,taneli} investigate
the problem whether there exists a data set that is consistent with
the given frequent itemsets and frequencies and show that this
problem is {\bf NP}-complete. The frequency of each frequent itemset
can be taken as a constraint over the original data set. The problem
of inverse frequent set mining then can be translated to a linear
constraint problem. Linear programming problems can be commonly
solved today in hundreds or thousands of variables and constraints.
However, the number of variables and constraints in this scenario is
far beyond hundreds or thousands (e.g., $2^t$, where $t$ is the
number of items). Hence it is impractical to apply linear
programming techniques directly.  Recently, the authors of
\cite{wwwl} investigated a heuristic method to generate synthetic
basket data set using the frequent sets and their supports mined
from the original basket data set. Instead of applying linear
programming directly on all the items, it applies graph-theoretical
results to decompose items into independent components and then
apply linear programming on each component. One potential problem
here is that the number of items contained in some components may be
still too large (especially when items are highly correlated each
other), which makes the application of linear programming
infeasible.

The authors of \cite{Ramesh:pods03,Ramesh:ideas05} proposed a method
to generate basket data set for benchmarking when the length
distributions of frequent and maximal frequent itemset collections
are available. Though the generated synthetic data set preserves the
length distributions of frequent patterns, one serious limitation is
that the size of transaction databases generated is much larger than
that of original database while the number of items generated is
much smaller. We believe the sizes of items and transactions are two
important parameters as they may significantly affect the
performance of association rule mining algorithms.

Instead of using the exact inverse frequent itemset mining approach,
we propose an approach to construct transaction databases
which have the same size as the original transaction database and which
are approximately consistent with the given frequent itemset
constraints. These approximate transaction databases are sufficient
for benchmarking purpose. In this paper, we consider the
complexity problem, the approximation problem, and privacy
issues for this approach.

We first introduce some terminologies.
${\cal I}$ is the finite set of items. A transaction over
${\cal I}$ is defined as a pair $(tid, I)$ where $I$ is a subset of
${\cal I}$ and tid is a natural number, called the transaction identifier.
A transaction database ${\cal D}$ over ${\cal I}$ is a finite set of
transactions over ${\cal I}$. For an item set $I\subseteq {\cal I}$
and a transaction $(tid, J)$, we say that  $(tid, J)$ contains $I$ if
$I\subseteq J$. The support of an itemset $I$ in a transaction database
${\cal D}$ over ${\cal I}$ is defined as the number of transactions $T$
in ${\cal D}$  that contains $I$, and is denoted $support(I, {\cal D})$.
The frequency of an itemset $I$ in a transaction database
${\cal D}$ over ${\cal I}$ is defined as
$$freq(I, {\cal D})=_{def}\frac{support(I, {\cal D})}{|{\cal D}|}.$$
Calders \cite{caldersthesis,calderspods} defined the following problems that
are related to the inverse frequent itemset mining.

\vskip 10pt
\noindent
FREQSAT

\noindent
{\it Instance}: An item set ${\cal I}$ and a sequence
$(I_1, f_1)$, $(I_2, f_2)$, $\cdots$, $(I_m,f_m)$,
where $I_i\subseteq {\cal I}$ are itemsets and $0\le f_i\le 1$
are nonnegative rational numbers, for all $0\le i\le m$.

\noindent
{\it Question}: Does there exist a transaction database ${\cal D}$ over
${\cal I}$ such that $freq(I_i, {\cal D})=f_i$ for all $0\le i\le m$?

\vskip 10pt
\noindent
FFREQSAT (Fixed size FREQSAT)
\nopagebreak

\noindent
{\it Instance}: An integer $n$, an item set ${\cal I}$, and a sequence
$(I_1, f_1)$, $(I_2, f_2)$, $\cdots$, $(I_m,f_m)$,
where $I_i\subseteq {\cal I}$ are itemsets and $0\le f_i\le 1$
are nonnegative rational numbers, for all $0\le i\le m$.

\noindent
{\it Question}: Does there exist a transaction database ${\cal D}$ over
${\cal I}$ such that ${\cal D}$ contains $n$ transactions and
$freq(I_i, {\cal D})=f_i$ for all $0\le i\le m$?

\vskip 10pt
\noindent
FSUPPSAT

\noindent
{\it Instance}: An integer $n$, an item set ${\cal I}$, and a sequence
$(I_1, s_1)$, $(I_2, s_2)$, $\cdots$, $(I_m,s_m)$,
where $I_i\subseteq {\cal I}$ are itemsets and $s_i\ge 0$ are nonnegative
integers, for all $0\le i\le m$.

\noindent
{\it Question}: Does there exist a transaction database ${\cal D}$ over
${\cal I}$ such that ${\cal D}$ contains $n$ transactions and
$support(I_i, {\cal D})=s_i$ for all $0\le i\le m$?

\vskip 5pt
Obviously, the problem FSUPPSAT is equivalent to the problem FFREQSAT.
Calders \cite{caldersthesis} showed that FREQSAT
is {\bf NP}-complete and the problem
FSUPPSAT is equivalent to the Intersection Pattern problem {\bf IP}:
given an $n\times n$ matrix $C$ with integer entries, do there exist
sets $S_1,\ldots, S_n$ such that $|S_i\cap S_j|=C[i,j]$?
Though it is known that {\bf IP} is {\bf NP}-hard, it is an open problem
whether {\bf IP} belongs to {\bf NP}.

In this paper, we will consider the problem of generating
transaction databases that approximately satisfy the given frequent
itemset support constraints. Section \ref{appcom} discusses the
computational complexity of approximating transaction databases.
Section \ref{appsection} proposes an algorithm to approximately
generate a approximate transaction database. Section
\ref{privacysec} discusses privacy issues and Section
\ref{relatedsec}. Finally, Section \ref{consection} draws
conclusions.

\section{Approximations}
\label{appcom}
Though it is an interesting problem to study whether there exists
a size $n$ transaction database that satisfies a set of
given frequency constraints, it is sufficient for benchmarking purpose
to construct a transaction database that is approximately at the size of $n$
and that approximately  satisfies the set of given frequency constraints.
Thus we define the following problem.

\vskip 10pt
\noindent
ApproSUPPSAT

\noindent
{\it Instance}: An integer $n$, an item set ${\cal I}$,
and a sequence $(I_1, s_1)$, $(I_2, s_2)$, $\cdots$, $(I_m,s_m)$,
where $I_i\subseteq {\cal I}$ are itemsets and $s_i\ge 0$ are nonnegative
integers, for all $0\le i\le m$.

\noindent
{\it Question}: Does there exist a transaction database ${\cal D}$
of $n'$ transactions over ${\cal I}$ such that $|n-n'|=O(m)$ and
$|support(I_i, {\cal D})-s_i|=O(m)$ for all $0\le i\le m$?

\vskip 5pt
Note that in the above definition, the approximation errors are
based on the parameter $m$ instead of $n$ since for most applications,
$m$ is small and $n$ is bigger. Indeed, $n$ could be at the exponential
order of $m$. For performance testing purpose, it is not meaningful
to use $n$ as the parameter in these situations. It also straightforward
to show that the problem ApproSUPPSAT is equivalent to the
following problem: given an integer $n$, an item set ${\cal I}$, and
a sequence $(I_1, s_1), (I_2, s_2), \cdots, (I_m,s_m)$,
decide whether there exists a transaction database ${\cal D}$ over
${\cal I}$ with $n$ transactions and
$0\le support(I_i, {\cal D})-s_i=O(m)$ for all $0\le i\le m$.

In the following we show that ApproSUPPSAT is
{\bf NP}-complete. Note that for the non-approximate version
FSUPPSAT of this problem, we do not know whether it is in {\bf NP}.

\begin{lemma}
\label{smallmodel}
ApproSUPPSAT $\in {\bf NP}$.
\end{lemma}

\noindent
{\bf Proof}. Since the size of the
transaction database is $n$ which might be exponential in the
size of the instance input description, it is not possible to guess
a transaction database in polynomial time and check whether it satisfies
the constraints. In the following, we use other
techniques to show that the problem is in {\bf NP}.
Let ${\cal I}$ be the collection of item sets and
$(I_1, s_1)$, $(I_2, s_2)$, $\cdots$, $(I_m,s_m)$ be the sequence of
support constraints. Assume that $|{\cal I}|=t$. Let
$J_0, J_1, \cdots, J_{2^t-1}$ be an enumeration of the $2^t$
subsets of ${\cal I}$ (in particular, let
$J_0=\emptyset$ and $J_{2^t-1}={\cal I}$), and
$X_0, X_1$, $\ldots$, $X_{2^t-1}$ be $2^t$ variables corresponding
to these itemsets.

Assume that a transaction database ${\cal D}$ with $n'=n+O(m)$ transactions
contains $X_i$ itemset $J_i$ for each $0\le i\le 2^t$ and  ${\cal D}$
approximately satisfies the support constraints
$(I_1, s_1)$, $(I_2, s_2)$, $\cdots$, $(I_m,s_m)$. Then there exists an
integer $k$ such that the following equations (\ref{leqa}) hold
for some integer values $X_0, \ldots, X_{2^t-1}$, $Z_0$, $\ldots$, $Z_m$.
Similarly, if there is an integer $k$ and an integer solution to
the equations (\ref{leqa}), then  there is a transaction database
${\cal D}$ with $n'=n+O(m)$ transactions that
approximately satisfies the support constraints $(I_1, s_1)$,
$\ldots$, $(I_m,s_m)$.
\begingroup
\def\arraystretch{1.3}
\begin{equation}
\label{leqa}
\begin{array}{ccc}
X_1, \ldots, X_{2^t}&\ge& 0\\
|Z_0|, |Z_1|, \ldots, |Z_m|&\le& km\\
\sum_{i=0}^{2^t}X_i +Z_0&=& n\\
\sum_{I_1\subseteq J_i}X_i+Z_1 &=& s_1\\
\cdots  &&\\
\sum_{I_m\subseteq J_i}X_i +Z_m&=& s_m
\end{array}
\end{equation}
\endgroup
where $k$ is a large enough integer.
In another word, if the given instance of the ApproSUPPSAT
problem is satisfiable, then the equations  (\ref{leqa}) have an integer
solution. That is, the solution space for the equation
(\ref{leqa}) is a non-empty convex polyhedron. A simple
argument\footnote{Similar argument has been used to prove the fundamental
theorem of linear optimization in linear programming. See, e.g.,
\cite{fagin,potts}.} could then
be used to show
that there is an extreme point $(X_1^0,\ldots, X_{2^t}^0)$ (not necessarily
an integer point) on this convex polyhedron that satisfies the
following property:
\begin{itemize}
\item There are at most $m+1$ non-zero values among
the variables $X^0_1$, $\ldots$, $X^0_{2^t}$, $Z_0$, $\ldots$, $Z_m$.
\end{itemize}
Let $Y_i=[ X^0_i]$ be the closest integer to $X^0_i$ for $1\le i\le 2^t$
and  ${\cal D}^Y$ be the transaction database that contains
$Y_i$ copies of the itemset $J_i$ for each $0\le i\le 2^t$. Then
${\cal D}^Y$ contains $n+O(m)$ transactions and
$|support(I_i, {\cal D})-s_i|=O(m)$ for all $0\le i\le m$.

In another word, the given instance of the ApproSUPPSAT
problem is satisfiable if and only if
there exist itemsets $J_1, \ldots, J_{m+1}$ and an integer sequence
$x_1, \ldots, x_{m+1}$ such that the transaction database
${\cal D}$ consisting of $x_i$ copies of itemset $J_i$ for each
$i\le m$ witnesses the satisfiability.
Thus ApproSUPPSAT$\ \in {\bf NP}$
which completes the proof of Lemma.
\hfill{Q.E.D.}

\begin{lemma}
\label{apphard}
ApproSUPPSAT is {\bf NP}-hard.
\end{lemma}

\noindent
{\bf Proof.}
The proof is based on an amplification of the reduction
in the {\bf NP}-hardness proof
for FREQSAT in \cite{caldersthesis} which is alike the one
given for 2SAT in \cite{psat}. In the following, we reduce the
{\bf NP}-complete problem 3-colorability to ApproSUPPSAT.
Given a graph $G=(V,E)$, $G$ is 3-colorable if there exists a 3-coloring
function $c:V\rightarrow \{R,G,B\}$ such that for each edge $(u,v)$
in $E$ we have $c(u)\not=c(v)$.

For the graph $G=(V,E)$, we construct an instance ${\cal A}(G)$ of
ApproSUPPSAT as follows. Let $m=6|V|+3|E|$, and $n=k_0m^2$
for some large $k_0$ (note that we need $k_0>k$ for the
constant $k$ we will discuss later).
Let the itemset $I=\{R_v,G_v,B_v: v\in B\}$ and the $m$ support
constraints are defined as follows. For each vertex $v\in V$:
\begingroup
\def\arraystretch{1.3}
$$\begin{array}{l}
support(\{R_v\})=[\frac{n}{3}], support(\{G_v\})=[\frac{n}{3}], \\
support(\{B_v\})=[\frac{n}{3}],\\
support(\{R_v,G_v\})=0, support(\{R_v,B_v\})=0,\\
support(\{G_v,B_v\})=0.
\end{array}$$
\endgroup
For each edge $(u,v)\in E$:
\begingroup
\def\arraystretch{1.3}
$$\begin{array}{l}
support(\{R_u,R_v\})=0, support(\{G_u,G_v\})=0,\\
support(\{B_u,B_v\})=0.\end{array}$$
\endgroup
In the following, we show that there is a transaction
database ${\cal D}$ satisfying this ApproSUPPSAT problem if and only
if $G$ is 3-colorable.

Suppose that $c$ is a 3-coloring of $G$. Let $T$ be a transaction
defined by letting $T_1=\{C_v:v\in V\}$ where
$$C_v=_{def}\left\{\begin{array}{ll}
R_v &\quad \mbox{ if } c(v)=R;\\
G_v &\quad  \mbox{ if } c(v)=G;\\
B_v &\quad  \mbox{ if } c(v)=B.\\
\end{array}\right.$$
Let transactions $T_2$ and $T_3$ be defined by colorings $c'$ and
$c''$ resulting from cyclically rearranging the colors $R,G,B$ in the
coloring $c$. Let the transaction database ${\cal D}$
consist of $[\frac{n}{3}]$ copies of each of the transaction
$T_1, T_2$, and $T_3$ (we may need to add one or two additional
copies of $T_1$ if $3[\frac{n}{3}]\not=n$).
Then ${\cal D}$ satisfies the ApproSUPPSAT problem
${\cal A}(G)$.

Suppose ${\cal D}$ is a transaction database  satisfying the
ApproSUPPSAT problem ${\cal A}(G)$. We will show that there is a
transaction $T$ in  ${\cal D}$ from which a 3-coloring of $G$ could
be constructed. Let ${\cal I}_1$ be the collection of itemsets
defined as
\begingroup
\def\arraystretch{1.3}
$$\begin{array}{l}
{\cal I}_1=\{\{R_v,G_v\}, \{R_v,B_v\}, \{G_v,B_v\}:v\in V\}\cup\\
\quad\quad\quad\quad
\{\{R_u,R_v\},\{G_u,G_v\},\{B_u,B_v\}: (u,v)\in E\}.\end{array}$$
\endgroup
That is, ${\cal I}_1$ is the collection of itemset that should
have $0$ support according to the support constraints.
Since ${\cal D}$ satisfies ${\cal A}(G)$,
for each $I'\in {\cal I}_1$, $support(I',{\cal D})=0$
is approximately satisfied. Thus there is
a constant $k_1>0$ such that at most $k_1m\times |{\cal I}_1|=3k_1m(|V|+|E|)$
transactions in  ${\cal D}$ contain an itemset in ${\cal I}_1$.
Let ${\cal D}_1$ be the transaction database obtained from
${\cal D}$ by deleting all transactions that contain
itemsets from ${\cal I}_1$. Then  ${\cal D}_1$ contains at least
$n-3k_1m(|V|+|E|)$ transactions.

For each vertex $v\in V$, we say that a transaction $(tid, J)$ in ${\cal D}$
does not contain $v$ if $J$ does not contain any items
from $\{R_v,G_v,B_v\}$. Since  ${\cal D}$ satisfies ${\cal A}(G)$,
for each $v\in V$, approximately one third of the
transactions contain $R_v$ ($G_v$, $B_v$, respectively). Thus there is
a constant $k_2>0$ such that at most $3k_2m\times |V|$ transactions
in  ${\cal D}$ do not contain some vertex $v\in V$. In another
word, there are at least $n-3k_2m\times |V|$ transactions $J$ in  ${\cal D}$
such that $J$ contains $v$ for all $v\in V$.

Let ${\cal D}_2$ be the transaction database obtained from
${\cal D}_1$ by deleting all transactions $J$ such that
$J$ does not contain some vertex $v\in V$. The above analysis shows
that  ${\cal D}_2$  contains at least
$n-3k_1m(|V|+|E|)- 3k_2m |V|$ transactions. Let
$k=\max\{k_1,k_2\}$. Then we have
\begingroup
\def\arraystretch{1.3}
$$\begin{array}{lll}
|{\cal D}_2| & \ge& n-3km(|V|+|E|)- 3km |V|\\
&=&n-km(6|V|+3|E|)\\
&=&n-km^2\\
&=&3\cdot k_0{m^2}-km^2
\end{array}$$
\endgroup
By the assumption of $k_0$ at the beginning of this proof,
we have $|{\cal D}_2|\ge 1$.
For any transaction $J$ in  ${\cal D}_2$, we can define a coloring
$c$ for $G$ by letting
$$c(v)=\left\{\begin{array}{ll}
R \quad\quad\mbox{ if } J \mbox{ contains } R_v\\
G \quad\quad\mbox{ if } J \mbox{ contains } G_v\\
B \quad\quad\mbox{ if } J \mbox{ contains } B_v\\
\end{array}\right.$$
By the definition of  ${\cal D}_2$, the coloring $c$ is defined
unambiguously. That is, $G$ is 3-colorable.

This completes the proof for {\bf NP}-hardness of ApproSUPPSAT.
\hfill{Q.E.D.}

\begin{theorem}
\label{npcompletetheorem}
ApproSUPPSAT is {\bf NP}-complete.
\end{theorem}

\noindent
{\bf Proof.} This follows from Lemma \ref{smallmodel}
and Lemma \ref{apphard}.
\hfill{Q.E.D.}

\vskip 5pt
We showed that the problem ApproSUPPSAT is {\bf NP}-hard.
In the proof of Lemma \ref{apphard}, we use the fact that
the number $n$ of transactions of the target basket database
is larger than the multiplication of the number $m$ of support
constraints and the approximate error $O(m)$
(that is, $n$ is in the order of $O(m^2)$).
In practice, the number $n$ may not be larger than
$km^2$. Then one may wonder whether the problem is still {\bf NP}-complete.
If $n$ is very small, for example, at the order of $O(m)$, then
obviously, the problem ApproSUPPSAT becomes trivial since one
can just construct the transaction database as the collection
of $n$ copies of the itemset ${\cal I}$ (that is, the entire set of items).
This is not a very interesting case since if $n$ is at the order
of $m$, one certainly does not want the approximate error to be at the
order of $n$ also. A reasonable problem could be that one defines
a constant number $\gamma$ to replace the approximate error $O(m)$.
Then the proof in Lemma \ref{apphard} shows that the
problem ApproSUPPSAT with approximate error $\gamma$ (instead of
$O(m)$)
is still {\bf NP}-complete if  $n>\gamma m$. Tighter bounds
could be achieved if weighted approximate errors for
different support constraints are given.

\section{Generating approximate transaction databases}
\label{appsection}
In this section, we design and analyze a linear
program based algorithm to approximate the {\bf NP}-complete
problem ApproSUPPSAT. Let ${\cal I}=\{e_1, \ldots, e_t\}$ be the
collection of items, $n$ be the number of transactions in the
desired database ${\cal D}$, and
$(I_1, s_1)$, $(I_2, s_2)$, $\cdots$, $(I_m,s_m)$ be the sequence of
support constraints.
According to the proof of Lemma \ref{smallmodel}, if this
instance of ApproSUPPSAT is solvable, then there is a
transaction database ${\cal D}$, consisting of at most
$m+1$ itemsets $J_1,\ldots, J_{m+1}$, that satisfies these constraints.
Let $X_1, \ldots, X_{m+1}$ be variables
representing the numbers of duplicated copies of these itemsets
in  ${\cal D}$ respectively. That is, ${\cal D}$ contains $X_i$ copies of
$J_i$ for each $i$.
For all $i\le m$ and $j\le m+1$, let $x_{i,j}$ and $y_{i,j}$
be variables with the property that $x_{i,j}=X_j\times y_{i,j}$ and
\begin{equation}
\label{LPc1}
y_{i,j}=\left\{
\begin{array}{ll}
1 & \mbox{ if } I_i\subseteq J_j,\\
0 & \mbox{ otherwise.}
\end{array}\right.
\end{equation}
Then we have $support(I_i, {\cal D})=x_{i,1}+\cdots+x_{i,m+1}$ and
the above given ApproSUPPSAT instance could be formulated
as the following question.
\begingroup
\def\arraystretch{1.3}
\begin{equation}
\mbox{minimize } z_1+z_2+\cdots + z_m
\label{targetLPcc}
\end{equation}
\endgroup
\nopagebreak
subject to
\nopagebreak
\begingroup
\def\arraystretch{1.3}
\begin{equation}
\left\{\begin{array}{l}
X_1+X_2+\cdots + X_{m+1}=n,\\
s_i+z_i=x_{i,1}+\cdots+x_{i,m+1}, \\
y_{i,j}=1 \mbox{ if } I_i\subseteq J_j \mbox{ and } y_{i,j}=0
        \mbox{ otherwise,}\\
x_{i,j}=X_j\times y_{i,j},\\
z_i,\ X_j \mbox{ are nonnegative integers},
\end{array}\right.
\label{LPc3}
\end{equation}
\endgroup
for $i\le m \mbox{ and } j\le m+1$.

The condition set (\ref{LPc3}) contains the nonlinear equation
$x_{i,j}=X_j\times y_{i,j}$ and the nonlinear condition
specified in (\ref{LPc1}).
Thus in order to approximate the given ApproSUPPSAT instance
using linear program techniques, we need to convert these conditions
to linear conditions.

We first use characteristic arrays of variables
to denote the unknown itemsets $J_1, \ldots, J_{m+1}$.
For any itemset $I\subseteq {\cal I}$, let the $t$-ary array
$\chi(I)\in\{0,1\}^t$ be the characteristic array of $I$. That is,
the $i$-th component $\chi(I)[i]=1$ if and only if $e_i\in I$.
Let $\chi({J_1})=(u_{1,1},\ldots, u_{1,t})$, $\ldots$,
$\chi({J_{m+1}})=(u_{m+1, 1},\ldots, u_{m+1,t})$ be a collection of
$(m+1)t$ variables taking values from $\{0,1\}$, representing
the characteristic arrays of $J_1,\ldots, J_{m+1}$ respectively.

In order to convert the condition specified in (\ref{LPc1})
to linear conditions. we first use inner product constraints to
represent the condition $I_i\subseteq J_j$.
For two characteristic arrays $\chi_1$ and $\chi_2$, their
inner product is defined as $\chi_1\cdot \chi_2=\chi_1[1]\cdot\chi_2[1]
+\cdots+\chi_1[t]\cdot\chi_2[t]$. It is straightforward to show that
for two itemsets $I,J\subseteq {\cal I}$, we have
$\chi(I)\cdot \chi(J)\le \min\{|I|,|J|\}$
and  $\chi(I)\cdot \chi(J)=|I|$ if and only if $I\subseteq J$.

\begin{center}
\begin{figure}[htb]
\begin{center}
{\includegraphics{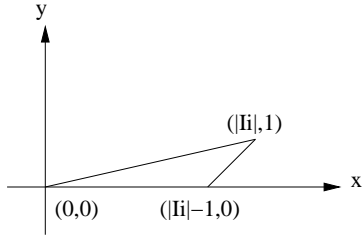}}
\end{center}
\caption{Triangle} \label{config1}
\end{figure}
\end{center}

Now the following conditions
in (\ref{LPeq1}) will guarantee that the condition in (\ref{LPc1})
is satisfied.
\begingroup
\def\arraystretch{1.3}
\begin{equation}
\left\{\begin{array}{l}
|I_i|\cdot y_{i,j}\le \chi({J_j})\cdot \chi({I_i})\le y_{i,j}+|I_i|-1\\
y_{i,j},\ u_{j,k}\in\{0,1\}
\end{array}\right.
\label{LPeq1}
\end{equation}
\endgroup
for all $i\le m$, $j\le m+1$, and $k\le t$.
The geometric interpretation of this condition is as follows.
If we consider  $(\chi({J_j})\cdot \chi({I_i}), y_{i,j})$
as a point in the 2-dimensional space $(x,y)$ shown in Figure \ref{config1},
then $|I_i|y\le x$ defines points below the line passing
the points $(0,0)$ and $(|I_i|,1)$, and $x\le y+|I_i|-1$ defines
the points above the line passing through the points
$(|I_i|-1,0)$ and $(|I_i|,1)$. Thus $y_{i,j}=1$ if and only
if $\chi({J_j})\cdot \chi({I_i})=|I_i|$.
That is, $y_{i,j}=1$ if and only if $I_i\subseteq J_j$.

\begin{center}
\begin{figure}[htb]
\begin{center}
{\includegraphics{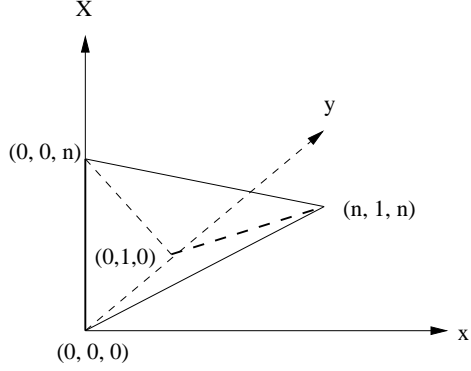}}
\end{center}
\caption{Tetrahedron} \label{fig1}
\end{figure}
\end{center}

The nonlinear equations $x_{i,j}=X_j\times y_{i,j}$
can be converted to the following conditions consisting of
inequalities.
\begingroup
\def\arraystretch{1.3}
\begin{equation}
\left\{\begin{array}{l}
x_{i,j}-ny_{i,j}\le 0,\\
X_j\ge x_{i,j}, \\
ny_{i,j}+X_j-x_{i,j}\le n, \\
x_{i,j}\ge 0, \\
y_{i,j}\in\{0,1\},
\end{array}\right.
\label{LPeq2}
\end{equation}
\endgroup
for all $i\le m$ and $j\le m+1$. The constant $n$ is used in the
inequalities due to the fact that $X_j\le n$ for all $j\le m+1$.
The geometric interpretation for the above inequalities is described in
the following. If we consider $(x_{i,j}, y_{i,j}, X_j)$ as a point
in a 3-dimensional space $(x,y,X)$ shown in Figure \ref{fig1}, then
\begin{enumerate}
\item $x-ny=0$ defines the plane passing through points
$(0,0,0)$, $(0,0,n)$, and $(n,1,n)$; Thus
$x_{i,j}-ny_{i,j}\le 0$ guarantees that $x_{i,j}=0$ if $y_{i,j}=0$.
\item \label{cond2}
$X\ge x$  defines the points above the plane
passing through points $(0,0,0)$, $(0,1,0)$, and $(n,1,n)$. This
condition together with the condition $y_{i,j}\in\{0,1\}$
guarantees that $x_{i,j}\le X_j$ when  $y_{i,j}=1$.
\item $ny+X-x\le n$ defines the points below the
plane passing through points $(0,1,0), (0,0,n)$, and $(n,1,n)$. This
condition together with the condition $y_{i,j}\in\{0,1\}$
guarantees that $x_{i,j}\ge X_j$ when  $y_{i,j}=1$.
Together with the condition \ref{cond2}, we have $x_{i,j}= X_j$
when  $y_{i,j}=1$.
\end{enumerate}

\noindent
{\bf Note}: For the reason of convenience, we introduced
the intermediate variables $y_{i,j}$. In order to improve
the linear program performance, we may combine the
conditions (\ref{LPeq1}) and (\ref{LPeq2}) to cancel the
variables $y_{i,j}$.

Thus the integer programming formulation for the given
ApproSUPPSAT instance is as follows.
\begingroup
\def\arraystretch{1.3}
\begin{equation}
\mbox{minimize } z_1+z_2+\cdots + z_m
\label{targetLP}
\end{equation}
\endgroup
\nopagebreak
subject to conditions (\ref{LPeq1}), (\ref{LPeq2}), and
\nopagebreak
\begingroup
\def\arraystretch{1.3}
\begin{equation}
\left\{\begin{array}{l}
X_1+X_2+\cdots + X_{m+1}=n,\\
s_i+z_i=x_{i,1}+\cdots+x_{i,m+1}, \\
z_i,\ X_j \mbox{ are nonnegative integers},
\end{array}\right.
\label{LPeq3}
\end{equation}
\endgroup
for $i\le m \mbox{ and } j\le m+1$.
We first solve the linear relaxation of this
integer program. That is, replace the second equation
in the condition (\ref{LPeq1}) by
$$0\le y_{i,j},u_{j,k}\le 1\quad\mbox{ for all }i\le m,
j\le m+1, \mbox{ and }k\le t$$
and replace the third equation in the condition (\ref{LPeq3}) by
$$z_i,\ X_j\ge 0.$$
Let $o^*=\{(u^*_{j,k}, y^*_{i,j}, x^*_{i,j}, z^*_i, X^*_j):
i\le m, j\le m+1, k\le t\}$
denote an optimal solution to this relaxed linear program.
There are several ways to construct an integer solution $\bar{o}$
from  $o^*$. Let $OPT(z;I)$ denote the optimal value of $z_1+\cdots+z_m$
for a given ApproSUPPSAT instance $I$ and $\overline{OPT}(z;I)$ be
the corresponding value for the computed integer solution.
For an approximation algorithm, one may prefer to compute
a number $\alpha$ such that
$$\overline{OPT}(z;I)\le \alpha OPT(z;I).$$
Theorem \ref{npcompletetheorem} shows that it is {\bf NP}-hard
to approximate the ApproSUPPSAT by an additive polynomial
factor. Thus  $\overline{OPT}(z;I)$ is not in the order
of $O(m)$ in the worst case for any polynomial time approximation
algorithms, and it is not very interesting to analyze the worst
case for our algorithm.

In the following, we first discuss two simple
naive rounding methods to get an integer solution $\bar{o}$
from  $o^*$. We then present two improved
randomized and derandomized rounding methods.

\subsection*{Method 1: rounding $u^*_{j,k}$}
Construct an integer solution
$\bar{o}=(\bar{u}_{j,k}, \bar{y}_{i,j}, \bar{x}_{i,j}, \bar{z}_i, \bar{X}_j)$
by rounding $u^*_{j,k}$ to their closest integers,
rounding $X^*_j$  to their almost closest integers
so that $\bar{X}_1+\cdots+\bar{X}_{m+1}=n$, and
computing $\bar{y}_{i,j}, \bar{x}_{i,j}$, and $\bar{z}_i$
according to their definitions.
That is, for each $j\le m+1$ and $k\le t$ set
$$\bar{u}_{j,k}=\left\{\begin{array}{ll}
1 \quad & \mbox{if }u^*_{j,k}\ge 0.5,\\\
0 \quad & \mbox{otherwise.}
\end{array}\right.$$
For the rounding of $X^*_j$, first round $X^*_{j}$ to their
closest integers $[X^*_j]$. Then randomly
add/subtract $1$'s to/from these values according to the value of
$\bar{X}_1+\cdots+\bar{X}_{m+1}-n$ until $\bar{X}_1+\cdots+\bar{X}_{m+1}=n$.

From the construction, it is clear that $\bar{o}$
is a feasible solution of the integer program. The rounding
procedure will introduce the following errors to the optimal solution:
\begin{enumerate}
\item By rounding $\{u^*_{j,k}:i\le m, k\le t\}$,
the values in $\{\chi(I_i)\cdot \chi(J_j):i\le m, j\le m+1\}$ change.
Thus the values in $\{\bar{y}_{i,j}:i\le m,j\le m+1\}$
will change. Thus the values in $\{\bar{x}_{i,j}:i\le m,j\le m+1\}$
will be different from the values in  $\{x^*_{i,j}:i\le m,j\le m+1\}$.
\item By rounding $\{X^*_j:j\le m+1\}$, the values of
$\{\bar{x}_{i,j}:i\le m,j\le m+1\}$ will change also.
\end{enumerate}

\subsection*{Method 2: rounding $x^*_{i,j}$}
Construct an integer solution
$\bar{o}=(\bar{u}_{j,k}, \bar{y}_{i,j}, \bar{x}_{i,j}, \bar{z}_i, \bar{X}_j)$
by rounding $x^*_{i,j}$ to $0$ or $X^*_j$ and
computing the other values according to their definitions or relationships.
That is, first round $X^*_{j}$ to their
closest integers $[X^*_j]$. Then randomly
add/subtract $1$'s to/from these values according to the value of
$\bar{X}_1+\cdots+\bar{X}_{m+1}-n$ until $\bar{X}_1+\cdots+\bar{X}_{m+1}=n$.
Now round  $x^*_{i,j}$ as follows. Let
$$\bar{x}_{i,j}=\left\{\begin{array}{ll}
\bar{X}_j \quad & \mbox{if }x^*_{i,j}\ge 0.5 \bar{X}_j,\\\
0 \quad & \mbox{otherwise.}
\end{array}\right.$$
$J_j$'s could be computed by setting
$$J_j=\cup_{\bar{x}_{i,j}=\bar{X}_j} I_i.$$
The values of $\bar{u}_{j,k}$ and $\bar{y}_{i,j}$ can be derived
from $J_j$ easily. We still need to further update the values of
$\bar{x}_{i,j}$ by using the current values of $\bar{y}_{i,j}$
since we need to satisfy the requirements $x_{i,j}=X_j\times y_{i,j}$.

From the construction, it is clear that $\bar{o}$
is a feasible solution of the integer program. The rounding
procedure will introduce the following errors to the optimal solution:
\begin{enumerate}
\item By rounding $\{x^*_{i,j}:i\le m, j\le m+1\}$,
we need to update the values of $\bar{y}_{i,j}$, which again
leads to the update of values of $\bar{x}_{i,j}$.
\item By rounding $\{X^*_j:j\le m+1\}$, the values in
$\{\bar{x}_{i,j}:i\le m,j\le m+1\}$ will change also.
\end{enumerate}

\subsection*{Method 3: randomized and derandomized rounding}
For quite a few {\bf NP}-hard problems that are reduced to
integer programs,
naive round methods remain to be the ones with best known
performance guarantee. Our methods 1 and 2 are based
on these naive rounding ideas. In last decades, randomization and
derandomization
methods (see, e.g., \cite{shmoys,raghavan}) have received a great deal of
attention in algorithm design.
In this paradigm for algorithm design, a randomized algorithm
is first designed, then the algorithm is ``derandomized'' by
simulating the role of the randomization in critical places
in the algorithm. In this section, we will design a randomized
and derandomized rounding approach to obtain an integer
solution $\bar{o}$ from $o^*$ with performance of at least the
expectation. It is done by the method of conditional probabilities.

In rounding method 1, we round $u^*_{j,k}$ to its closest integer.
In a random rounding \cite{randomrounding}, we set the value of
$\bar{u}_{j,k}$ to  $1$ with probability $u^*_{j,k}$ and to $0$ with
probability $1-u^*_{j,k}$ (independent of other indices).

In rounding method 2, we round $x^*_{i,j}$ to the closest value
among $0$ and $\bar{X}_j$. In a random rounding \cite{randomrounding},
we set the value of $\bar{x}_{i,j}$ to  $\bar{X}_j$
with probability $\frac{x^*_{i,j}}{\bar{X}_j}$ and to $0$ with
probability $1-\frac{x^*_{i,j}}{\bar{X}_j}$ (independent of
other indices).

A random rounding approach produces integer
solutions with an expected value $z_0$ for $\sum_{i=1}^mz_i$.
An improved rounding approach (derandomized rounding)
produces integer solutions with
$\sum_{i=1}^mz_i$ guaranteed to be no larger than the expected value $z_0$.
In the following, we illustrate our method for the random
rounding based on the rounding methods 1 and 2.

\noindent
{\bf Randomized and derandomized rounding of $x^*_{i,j}$.}
We determine the value of an additional variable in each step.
Suppose that $\{\bar{x}_{i,j}: (i,j)\in I_0\}$ has already
been determined, and we want to determine the value of
$\bar{x}_{i_0,j_0}$ with $(i_0,j_0)\notin I_0$.
We compute the conditional expectation for  $\sum_{i=1}^mz_i$
of this partial assignment first with $\bar{x}_{i_0,j_0}$
set to zero, and then again with it set to $\bar{X}_{j_0}$.
If we set  $\bar{x}_{i_0,j_0}$ according to which of these values is
smaller, then the conditional expectation at the end of this step
is at most the conditional expectation at the end
of the previous step. This implies that at the end of the
rounding, we get at most the original expectation.

In the following, we show how to compute the conditional expectation.
At the beginning of each step, assume that for all entries
$(i',j')$ in $I_0$, $\bar{x}_{i',j'}$ has been determined already
and we want to determine the value of $\bar{x}_{i_0,j_0}$ for
$(i_0,j_0)\notin I_0$ in this step.

In order to compute
the conditional expectation of $\sum_{i=1}^mz_i$, we first
compute the probability $\mbox{Prob}[I_i\subseteq J_j]$
for all $(i,j)\notin I_0$. For each $j\le m+1$, let
$$J^0_j=\displaystyle\bigcup_{I_{i'}\subseteq J_{j}, (i',j)\in I_0} I_{i'}$$
If $I_i\subseteq J^0_j$, then we have $I_i\subset J_j$
and $\mbox{Prob}[I_i\subseteq J_j]=1$.  Otherwise, continue with the
following computation.
By regarding $\frac{x^*_{i,j}}{\bar{X}_j}$ as the
probability that $x^*_{i,j}$ takes the value ${\bar{X}_j}$, we know that
with at least probability  $\frac{x^*_{i,j}}{\bar{X}_j}$
we have $I_i\subseteq J_j$. However, the actual probability may be larger
since other entries $I_{i'}$ with $I_i\cap I_{i'}\not=\emptyset$
may contribute items to $J_j$, which may lead to the inclusion of
$I_i$ in $J_j$. First we define the following sets.
\begingroup
\def\arraystretch{1.3}
$$\begin{array}{lll}
L_{i,j} & = & \{1, \ldots, {i-1}, {i+1},\ldots m\}\setminus\{i': (i',j)\in
I_0\}\\
U_{i,j}&=&\left\{K\subseteq L_{i,j}:
I_i\subseteq J^0_j\bigcup \displaystyle\bigcup_{i'\in K}I_{i'}\right\},
\end{array}$$
\endgroup
and
$$U'_{i,j}=\{K\in U_{i,j}: \mbox{there is no }K'\in U_{i,j}
\mbox{ such that }K'\subset K \}.$$
For each $K\in U'_{i,j}$, let
$$p(i,j,K)=\prod_{i'\in K}\frac{x^*_{i',j}}{\bar{X}_j}.$$
Then the probability $\mbox{Prob}[I_i\subseteq J_j]$ can be approximated as
$$\mbox{Prob}[I_i\subseteq J_j]=
\frac{x^*_{i,j}}{\bar{X}_j} + \left(1-\frac{x^*_{i,j}}{\bar{X}_j}\right)
\sum_{K\in U'_{i,j}} p(i,j,K).$$
Note that we say that we approximate the probability
$\mbox{Prob}[I_i\subseteq J_j]$ since in the computation,
we assume that $\mbox{Prob}[I_{i'}\subseteq J_{j}]=
\frac{x^*_{i',j}}{\bar{X}_j}$ for other $i'$ which may not
be true. If necessary,
we can improve the approximation by iteration. That is, repeat
the above procedure for several rounds and, in each round,
use the approximated probabilities for
$\mbox{Prob}[I_{i'}\subseteq J_{j}]$
from the previous round. If sufficient rounds are repeated,
the probability will converge in the end.

Since we have the probabilities $\mbox{Prob}[I_i\subseteq J_j]$ for all
$(i,j)\notin I_0$ now, it is straightforward to compute the
conditional expectation of  $E(\sum_{i=1}^mz_i)=
\sum_{i=1}^mE(z_i)$. The expected
value for $z_i$ is
$$E(z_i)=E\left(\sum_{j=1}^{m+1}x_{i,j}\right)-s_i=
\sum_{j=1}^{m+1}\bar{X}_j\cdot\mbox{Prob}[I_i\subseteq J_j] - s_i.$$

\noindent
{\bf Randomized and derandomized rounding of $u^*_{j,k}$.}
We determine the value of an additional variable in each step.
Suppose that $\{\bar{u}_{j,k}: (j,k)\in I_0\}$ has already
been determined, and we want to determine the value of
$\bar{u}_{j_0,k_0}$ with $(j_0,k_0)\notin I_0$.
We compute the conditional expectation for  $\sum_{i=1}^mz_i$
of this partial assignment first with $\bar{u}_{i_0,j_0}$
set to zero, and then again with it set to $1$.
If we set  $\bar{u}_{j_0,k_0}$ according to which of these values is
smaller, then the conditional expectation at the end of this step
is at most the conditional expectation at the end
of the previous step. This implies that at the end of the
rounding, we get at most the original expectation.

According to our analysis in the randomized and derandomized
rounding of $x^*_{i,j}$, it is sufficient to compute the
probability $\mbox{Prob}[I_i\subseteq J_j]$ for all $(i,j)$.
Assume ${\cal I}=\{e_1, \ldots, e_t\}$ and
$I_i=\{e_{i_1}$, $\ldots$, $e_{i_{|I_i|}}\}$.
Set
$$\mbox{Prob}[I_i\subseteq J_j]=
\hat{u}_{j, i_1}\times\cdots\times\hat{u}_{j, i_{|I_i|}}$$
where
$$\hat{u}_{j, i_s}=\left\{
\begin{array}{ll}
\bar{u}_{j, i_s} & \quad \mbox{ if } (j, i_s)\in I_0,\\
u^*_{j, i_s}     & \quad \mbox{ otherwise}
\end{array}\right.$$
for $s\le |I_i|$. Using $\mbox{Prob}[I_i\subseteq J_j]$,
one can compute the conditional expectation of $\sum_{i=1}^mz_i$
as in the case for rounding of $x^*_{i,j}$.

\subsection*{Complexity analysis of the approximation algorithm}
In the integer linear program formulation of our problem, we have
$t(m+1)$ variables $u_{j,k}$, \
$m+1$ variables $X_j$, \
$m(m+1)$ variables $x_{i,j}$, \
$m(m+1)$ variables $y_{i,j}$, and
$m$ variables $z_i$.
In total, we have $t(m+1)+2m^2+4m+1$ variables.

There are $(m+1)(2m+t)$ constraints in
the condition (\ref{LPeq1}), $4m(m+1)$ constraints in
the condition (\ref{LPeq1}), and $3m+2$ constraints in the
condition (\ref{LPeq3}). Thus we have $6m^2+9m+mt+t+2$ constraints
in total.

The rounding, randomized, and derandomized rounding algorithms
could be finished in $O(tm^3)$ steps. Thus the major challenge
is to solve the relaxed continuous variables linear program.
According to \cite{LPfaq}, hundreds of thousands of continuous
variables are regularly solved. Thus our approximation algorithm
are efficient when $m$ and $t$ takes reasonable values.

\section{Privacy issues}
\label{privacysec}
Wang, Wu, and Zheng \cite{wwz} considered general
information disclosure in the process of mock database generation.
In this section, we discuss privacy disclosures
in synthetic transaction databases.
Confidential information in transaction databases may be
specified as a collection of itemsets and their
corresponding support (frequency) intervals.
Let ${\cal P}$ be a set defined as follows.
$${\cal P}=\{(I_i, s_i, S_i): I_i\subseteq {\cal I}, i\le l\}.$$
We say that a (synthetic) transaction database  ${\cal D}$
does not disclose confidential information specified in
${\cal P}$ if one cannot infer that
$$s_i\le support(I_i;{\cal D})\le S_i$$
for all $(I_i, s_i, S_i)\in {\cal P}$.
Similarly, we say that a support constraint set
${\cal S}=\{(I'_1,s_1), \ldots, (I'_m,s_m)\}$
does not disclose confidential information specified in
${\cal P}$ if for each element $(I_i, s_i, S_i)\in {\cal P}$,
there is a transaction database ${\cal D}_i$
that satisfies all support constraints in ${\cal S}$ and
$$support(I_i,{\cal D}_i)\notin [s_i,S_i].$$

For the synthetic transaction database generation,
there are two scenarios for potential private information disclosure.
In the first scenario, the database owner uses the following
procedure to generate the synthetic transaction database:
\begin{enumerate}
\item use a software package to mine the real-world transaction
database to get a set of itemset support (frequency) constraints;
\item use a software package based on our linear program methods
to generate a synthetic transaction database ${\cal D}$ from the support
(frequency) constraints;
\item release the synthetic transaction database  ${\cal D}$ to the public.
\end{enumerate}
In this scenario, the mined support (frequency) constraints are
not released to the public and only the synthetic transaction
database is released.
In this case, it is straightforward to protect the confidential information
specified in ${\cal P}$. The database owner proceeds
according to the above steps until step $3$. Before releasing the
synthetic transaction database  ${\cal D}$, he can delete the confidential
information as follows.
\begin{itemize}
\item For each $(I_i, s_i, S_i)\in{\cal P}$, chooses
a random number $r_i\le n$, where $n$ is the total number
of transactions. We distinguish the following two cases:
\begin{enumerate}
\item If $u_i=support(I_i, {\cal D})-r_i< 0$, then
chooses a random series of $-u_i$ transactions $t_j$ that do not
contain the itemset $I_i$, and modify these transactions to contain
the itemset $I_i$.
\item If $u_i=support(I_i, {\cal D})-r_i\ge 0$, then
chooses a random series of $u_i$ transactions $t_j$ that
contain the itemset $I_i$, and modify these transactions in a random
way so that they do not contain the itemset $I_i$.
\end{enumerate}
\end{itemize}
After the above process, the resulting transaction database
contains no confidential information specified in ${\cal P}$
and the database owner is ready to release it.

In the second scenario, the database owner uses the following
procedure to generate the synthetic transaction database:
\begin{enumerate}
\item use a software package to mine the real-world transaction
database to get a set of itemset support (frequency) constraints;
\item release the support (frequency) constraints to the public;
\item a customer who has interest in a synthetic transaction database
generates a synthetic transaction database ${\cal D}$ from the
published support (frequency) constraints using a software package based on
our linear program methods.
\end{enumerate}
In this scenario, the mined support (frequency) constraints are
released to the public directly. Thus the database owner wants to
make sure that no confidential information specified in ${\cal P}$
is contained in these support (frequency) constraints.
Without loss of generality, we assume that there is a single
element $(I, s, S)$ in ${\cal P}$  and the mined
support constraints are ${\cal S}=\{(I_i,s_i): i\le m\}$.
${\cal S}$ contains the confidential information
$(I, s, S)$ if and only if for each transaction database ${\cal D}$
which is consistent with ${\cal S}$, we have
$support(I;{\cal D})\in [s, S]$. In another word,
${\cal S}$ does not contain the confidential information
$(I, s, S)$ if and only if there exists an integer
$s'$ with $s'<s$ or $S<s'<n$ such that
${\cal S}\cup \{(I,s')\}$ is consistent. That is,
there is a transaction database ${\cal D}$ that satisfies
all support constraints in ${\cal S}\cup \{(I,s')\}$.
In the following, we show that there is even no
efficient way to approximately decide whether
a given support constraint set contains confidential information.
We first define the problem formally.

\vskip 10pt
\noindent
ApproPrivacy

\noindent
{\it Instance}: An integer $n$, an item set ${\cal I}$,
a support constraint set
${\cal S}=\{(I'_1, s'_1)$, $\cdots$, $(I'_m,s'_m)\}$,
and a set ${\cal P}=\{(I_i, s_i, S_i): I_i\subseteq {\cal I}, i\le l\}.$

\noindent
{\it Question}: For all transaction database ${\cal D}$
of $n$ transactions over ${\cal I}$ with
$|support(I'_i, {\cal D})-s'_i|=O(m)$ for all $0\le i\le m$,
do we have $support(I_i, {\cal D})\in [s_i,S_i]$ for all
$i\le l$? If the answer is yes, we write ${\cal S}\models_{a} {\cal P}$.

\vskip 5pt

By Theorem \ref{npcompletetheorem}, we have the following
result. Similar {\bf NP}-hardness results for exact frequency
constraints inference have been obtained
in \cite{caldersthesis,calderspods,taneli}.
\begin{theorem}
\label{priNP}
ApproPrivacy is co{\bf NP}-complete.
\end{theorem}

{\bf Proof.} ${\cal S}\not\models_{a} {\cal P}$
if and only if there is a transaction database ${\cal D}$ and
an index $j\le l$ such that  ${\cal D}$ satisfies ${\cal S}\cup
\{(I_j, support(I_j, {\cal D})< s_i)\}$
or  ${\cal D}$ satisfies ${\cal S}\cup
\{(I_j, support(I_j, {\cal D})> S_i)\}$ approximately. Thus
the theorem follows from Theorem \ref{npcompletetheorem}.
\hfill{Q.E.D.}

Thus there is no efficient way for the database owner
to decide whether a support constraint set ${\cal S}$
leaks confidential information specified in ${\cal P}$.
In practice, however,
we can use the linear program based approximation
algorithms that we have discussed in Section \ref{appsection}
to compute the confidence level about private information
leakage as follows.
\begin{enumerate}
\item Convert the condition
${\cal S}\cup \{(I,s'): s'<s \mbox{ or } S<s'\le n\}$
to an integer linear program in the format of (\ref{LPeq3}). Note
that the condition  ``$s'<s \mbox{ or } S<s'\le n$''
is equivalent to the existential clause
$\exists s' \left((s'<s) \vee (S<s'\le n)\right)$. Thus it is
straightforward to convert it to integer linear program conditions.
\item Let the confidence level be  $c=\sum_{i=1}^mz_i$.
The smaller $c$, the higher confidence. In the ideal case
of $c=0$, we have found an itemset transaction database
${\cal D}$ that witnesses that no confidential information
specified by $(I,s,S)$ is leaked in ${\cal S}$.
\end{enumerate}
If the database owner thinks that the confidence
value $c=\sum_{i=1}^mz_i$ obtained in the above procedure
is too larger (thus confidence level is too low).
He may use the following procedure to delete potential
confidential information from the support constraint set.
\begin{enumerate}
\item Let $i$ be the number that maximizes
$\max_{(I_i,s_i)\in{\cal S}}|I\cap I_i|$.
\item Modify the value $s_i$ to be a random value.
\item Approximately revise support constraint values in ${\cal S}$
to make it consistent. For example, to make it satisfy the
monotonic rule. Since it is {\bf NP}-hard to determine whether
a support constraint set is consistent, we can only
revise the set  ${\cal S}$ to be approximately consistent.
\end{enumerate}
It should be noted that after the above process, the resulting
support constraint set may become inconsistent. Thus in the next
round, the value  $c=\sum_{i=1}^mz_i$ may be larger.
If that happens, the larger value $c$ does not interpret as the
privacy confidence level. Instead, it should be interpreted
as an indicator for inconsistency of the support constraint
set. Thus the above privacy deletion procedure should only be
carried out one time.

We should note that even if the confidence level is higher,
(that is,  $c=\sum_{i=1}^mz_i$  is small), there is still possibility
that the confidential information specified by $(I,s,S)$ is leaked
in theory. That is, for each transaction database ${\cal D}$
that satisfies the constraints  ${\cal S}$, we have
$support(I,{\cal D})\in [s,S]$. However, no one
may be able to recover this information since it is {\bf NP}-hard
to infer this fact. Support constraint inference has been extensively
studied by Calders in \cite{caldersthesis,calderspods}.

It would be interesting to consider conditional privacy-preserving
synthetic transaction database generations. That is, we say
that no private information is leaked unless some hardness problems are
solved efficiently. This is similar to the methodologies
that are used in public key cryptography. For example,
we believe that RSA encryption scheme is secure unless one
can factorize large integers.

In our case, we may assume that it is hard on average to efficiently
solve integer linear programs. Based on this assumption, we can say
that unless integer linear programs could be solved efficiently on average,
no privacy specified in ${\cal P}$ is leaked by ${\cal S}$
if the computed confidence level $c=\sum_{i=1}^mz_i$ is small.

\section{Related Work}
\label{relatedsec}

Privacy preserving data mining has been a very active research topic
in the last few years. There are two general approaches mainly from
privacy preserving data mining framework: data perturbation and the
distributed secure multi-party computation approach. As the context
of this paper focuses on data perturbation for single site, we will
not discuss the multi-party computation based approach for
distributed cases (See \cite{Pinkas:kddexp02} for a recent survey).

Agrawal and Srikant, in \cite{AS00}, first proposed the development
of data mining techniques that incorporate privacy concerns and
illustrated a perturbation based approach for decision tree
learning. Agrawal and Agrawal, in \cite{aa}, have provided a
expectation-maximization (EM) algorithm for reconstructing the
distribution of the original data from perturbed observations.  They
provide information theoretic measures to quantify the amount of
privacy provided by a randomization approach. Recently, Huang et al.
in \cite{Du:sigmod05}, investigated how correlations among
attributes affect the privacy of a data set disguised via the random
perturbation scheme and proposed methods (PCA based and MLE based)
to reconstruct original data. The objective of all randomized based
privacy-preserving data mining
\cite{aa,AS00,ESA+02,kargupta:icdm03,RH02} is to prevent the
disclosure of confidential individual values while preserving
general patterns and rules. The idea of these randomization based
approaches is that the distorted data, together with the
distribution of the random data used to distort the data, can be
used to generate an approximation to the original data values while
the distorted data does not reveal private information, and thus is
{\em safe} to use for mining. Although privacy preserving data
mining considers seriously how much information can be inferred or
computed from large data made available through data mining
algorithms and looks for ways to minimize the leakage of
information, however, the problem how to quantify and evaluate the
tradeoffs between data mining accuracy and privacy is still open
\cite{Evfimievski:pods03}.

In the context of privacy preserving association rule mining, there
have also been  a lot of active researches. In
\cite{Atallah:kdex99,Dasseni:ihw01}, the authors considered the
problem of limiting disclosure of sensitive rules, aiming at
selectively hiding some frequent itemsets from large databases with
as little impact on other, non-sensitive frequent itemsets as
possible. The idea was to modify a given database so that the
support of a given set of sensitive rules decreases below the
minimum support value. Similarly, the authors in
\cite{Saygin:sigmodr01} presented a method for selectively replacing
individual values with unknowns from a database to prevent the
discovery of a set of rules, while minimizing the side effects on
non-sensitive rules. The authors studied the impact of hiding
strategies in the original data set by quantifying how much
information is preserved after sanitizing a data set
\cite{Oliveira:icdm03}.  The authors, in \cite{ESA+02,RH02}, studied
the problem of mining association rules from transactions in which
the data has been randomized to preserve privacy of individual
transactions. One problem is it may introduce some false association
rules. The authors, in \cite{KC02,VC02}, investigated distributed
privacy preserving association rule mining. Though this approach can
fully preserve privacy, it works only for distributed environment
and needs sophisticated protocols (secure multi-party computation
based \cite{Yao86}), which makes it infeasible for our scenario.

Wu et al. have proposed a general framework for privacy preserving
database application testing by generating synthetic data sets based
on some a-priori knowledge about the production databases
\cite{wwz}. The general a-priori knowledge such as statistics and
rules can also be taken as constraints of the underlying data
records. The problem investigated in this paper can be thought as a
simplified problem where data set here is binary one and constraints
are frequencies of given frequent itemsets. However, the techniques
developed in \cite{wwz} are infeasible here as the number of items
are much larger than the number of attributes in general data sets.

\section{Conclusions}
\label{consection}
In this paper, we discussed the general problems
regarding privacy preserving synthetic transaction database
generation for benchmark testing purpose. In particular, we showed
that this problem is generally {\bf NP}-hard. Approximation
algorithms for both synthetic transaction database generation and
privacy leakage confidence level approximation have been proposed.
These approximation algorithms include solving a continuous variable
linear program. According to \cite{LPfaq}, linear problems having
hundreds of thousands of continuous variables are regularly solved.
Thus if the support constraint set size is in the order of hundreds
of thousands, then these approximation algorithms are efficient on
regular Pentium-based computers. If more constraints are necessary,
then more powerful computers are needed to generate synthetic
transaction databases.

\end{document}